\documentstyle[emulateapj,apjfonts,epsf]{article}

\newcommand{\Msun}{$M_{\odot}$}

\def\h2{H$_{2}$}

\input epsf

\begin{document}

\title{COSMOLOGICAL EFFECTS OF THE FIRST STARS:
EVOLVING SPECTRA OF POPULATION III}
\author{JASON~TUMLINSON\altaffilmark{1}, J. MICHAEL SHULL\altaffilmark{2}, \&
APARNA VENKATESAN\altaffilmark{3}} \affil{Center for Astrophysics and
Space
Astronomy, Department of Astrophysical and Planetary Sciences,\\
University of Colorado, Boulder, CO 80309} \altaffiltext{1}{Present
address: Department of Astronomy and Astrophysics, University of
Chicago, Chicago, IL, 60637}\altaffiltext{2}{Also at JILA, University
of Colorado and National Institute of Standards and
Technology}\altaffiltext{3}{NSF Astronomy and Astrophysics Postdoctoral
Fellow}

\begin{abstract}
The first stars hold intrinsic interest for their uniqueness and for
their potentially important contributions to galaxy formation, chemical
enrichment, and feedback on the intergalactic medium (IGM). Although
the sources of cosmological reionization are unknown at present, the
declining population of large bright quasars at redshifts $z > 3$
implies that stars are the leading candidates for the sources that
reionized the hydrogen in the IGM by $z \sim 6$. The metal-free
composition of the first stars restricts the stellar energy source to
proton-proton burning rather than the more efficient CNO cycle.
Consequently they are hotter, smaller, and have harder spectra than
their present-day counterparts of finite metallicity. We present new
results from a continuing study of metal-free stars from a cosmological
point of view. We have calculated evolving spectra of Pop III clusters,
derived from a grid of zero-metallicity stellar evolutionary tracks. We
find that H-ionizing photon production from metal-free stellar clusters
takes twice as long as that of Pop II to decline to 1/10 its peak
value. In addition, metal-free stars produce substantially more
photons than Pop II in the He~II ($E > 4$ Ryd) continuum. We suggest
that large Ly$\alpha$ equivalent widths ($W_{\rm Ly\alpha}
> 400$ \AA) may provide a means of detecting metal-free stellar
populations at high redshift, and that He~II recombination lines
($\lambda1640$, $\lambda4686$) may confirm identifications of
Population III. While Pop III clusters are intrinsically bluer
than their Pop II counterparts, nebular continuum emission makes
up this difference and may confuse attempts to discern Pop III
stars with broadband colors. In a companion paper, we explore the
consequences of evolving spectra of Pop III for the reionization
of the IGM in both H and He.
\end{abstract}

\section{Introduction}

The first stars and their role in galaxy formation and the evolution of
the intergalactic medium (IGM) define one of the frontiers of modern
cosmology. Yet we do not know when the first stars formed, where they
now reside, what happened to them over time, or even where to look for
them. The Big Bang paradigm requires that the first stars were composed
of H and He with only trace light elements and negligible carbon
(Schramm \& Turner 1998). Metal-free stars have unique properties owing
to this primordial composition: they are hotter and smaller than their
metal-enriched counterparts (Tumlinson \& Shull 2000, hereafter TS).
These features have interesting implications for their formation, their
environment, and particularly for their spectra. Some key open issues
are the formation epoch of the first stars (Ricotti, Gnedin, \& Shull
2002), the mass distribution for stars formed from primeval gas (Bromm,
Coppi, \& Larson 2001), and their contribution to the reionization of H
and He in the IGM. The recent suggestions by Lanzetta et al. (2002)
that star formation may become more widespread in the distant past, the
discovery of a gravitationally-lensed galaxy at $z = 6.56$ (Hu et al.
2002), and the apparent completion of H reionization by $z \simeq 6$
(Becker et al. 2001; Djorgovski et al. 2001) and He at $z \simeq 3$
(Kriss et al. 2001) make detailed examination of the first stars
particularly timely.

With our present tools we can see back to $z \sim 6$, roughly 1
Gyr after the Big Bang (a universe with $\Omega_m$ =  0.3,
$\Omega_\Lambda$ = 0.7, and $H_0$ = 70 km s$^{-1}$ Mpc$^{-1}$ is
assumed here). Imaging and spectroscopic surveys that sample wide
swaths of sky for telltale emission lines (Ly$\alpha$ emitters; Hu
et al.~1999; Rhoads et al.~2000) or peculiar broadband colors
(Lyman-break galaxies, Steidel et al. 1999) have discovered more
than 1000 galaxies at $z \sim 3$. Searches such as the Sloan
Digital Sky Survey (SDSS) are expanding the list of known QSOs at
$z \gtrsim 5$ and enabling high-redshift probes of the IGM (Fan et
al.~2001; Schneider et al.~2002). Ground-based spectra can
determine the star formation rate, metallicity, and mass of
galaxies at $z = 3$ (Pettini et al.~2002). The first stars are
believed to have formed between $z = 20 - 30$ (Gnedin \& Ostriker
1997; Ricotti et al. 2002), while galaxies have been discovered up
to $z \sim 6$. The next observational frontier is the direct
observation of the first stellar generation.

This work is part of an ongoing effort to examine the first stars from
a unified stellar and cosmological point of view. We began with
zero-age models (TS) of metal-free stars that revealed their potential
importance to H and He reionization in the IGM. We followed that with a
detailed calculation of the He~II recombination line fluxes from
Population III clusters, assessing the prospects for discovering this
emission as a distinctive signature of the first stars (Tumlinson,
Giroux, \& Shull 2001, hereafter TGS). We are concerned here with the
properties of the first stars that affect their observational
signatures and feedback on the IGM. These effects depend on the answer
to a fundamental question about the first stars: what is the time
evolution of their structure and radiation? We update the work of TS to
compute new evolving spectra of metal-free stars and to assess the
effects of evolution on the emergent spectra of Population III
clusters.

In this paper, we use the terms ``Population III'' and
``metal-free'' interchangeably. We also note that there are two
distinct usages of the term ``Pop III'' that have appeared in the
literature. One usage connotes small protogalactic clusters that
have $M_{\rm DM} \simeq 10^{6-7}$ \Msun\ and that must cool with
molecular hydrogen lines in the absence of metals. We adhere to
the stellar definition, in which ``Pop III'' and metal-free stars
are identical regardless of location.

In \S~2 we discuss the major components of our stellar evolution
calculations. In \S~3 we present the stellar evolution tracks that
enable our evolving spectra and observational predictions. In \S~4
we describe the modeling of the stellar atmospheres and construct
evolving spectra for synthetic Pop III clusters. In \S~5 we make
detailed predictions of the observational signatures of metal-free
stars and describe the realistic prospects for their discovery. In
\S~6 we draw some general conclusions. Paper II (Venkatesan,
Tumlinson, \& Shull 2002) presents the results of a cosmological
reionization model that incorporates Pop III spectra as ionizing
sources.

\section{Stellar Evolution at Zero Metallicity}

\subsection{Method}

To study the evolution of Population III stars, we must construct
realistic numerical models that account for the physical processes that
are thought to be relevant in real stars. The basic theory of stellar
structure is quite well developed and has been tested against
observations for decades. Occasional conflicts with detailed
observations lead to refinements of the theory, usually in the form of
additional input physics (convective overshoot, mass loss, and
rotation, for example). Of course, no such fine calibration of models
for Pop III stars is yet possible, since we have not discovered any of
these stars or anything believed to be their remnants. We must make the
basic assumption that the framework of stellar evolution theory applies
equally well to the unusual regime of zero metallicity. Under this
assumption, our Population III models incorporate the same paradigmatic
assumptions and physical ingredients that underlie all modern stellar
evolution codes. We must note here, however, that since our ultimate
goal is a calculation of the evolving spectra, we are not concerned
with internal processes, particularly late in the stellar lifetime,
that do not affect the spectral evolution. Those readers who are
interested in the details of the later stages of Pop III evolution are
referred to the tracks by Marigo et al. (2000) and Siess, Livio, \&
Lattanzio (2002). Schaerer (2002) has compiled tracks from a variety of
sources (El Eid et al. 1983; Marigo et al. 2000; Klapp 1983) in an
investigation into the properties of metal-free stars that parallels
some of our own results. Finally, we note that the following exposition
of our method is included for readers interested in the details. The
basic results of our study can be found by skipping to \S~3.

In our code we solve the fundamental stellar structure equations:
\begin{eqnarray}
 \frac{dR}{dM_r}   & = & \frac{1}{4 \pi R^2\rho} \\
 \frac{dL_r}{dM_r} & = & \epsilon - T\frac{dS}{dt}\\
 \frac{dP}{dM_r}   & = & - \frac{GM_r}{4\pi R^4}  \\
 \frac{dT}{dM_r}   & = & - \frac{GM_r T}{4 \pi R^4 P} \nabla,
\end{eqnarray}
where radius $R$, luminosity $L_r$, pressure $P$, and temperature $T$
are the ``primary variables'' and the subscript $r$ implies that the
variable represents a cumulative quantity up to radius $R$. These
equations are transformed to a Lagrangian mass grid that does not
change with total stellar mass. The mass density $\rho$, opacity
$\kappa$, energy generation rate $\epsilon$, thermodynamic derivatives
$\nabla_{\rm rad}$ and $\nabla_{\rm ad}$, and the specific entropy $S$
are ``secondary variables'' which are themselves functions of $T$, $P$,
$L$, $R$, and the chemical abundances by number, denoted by $X_i$. For
radiative energy transport, we use: $ \nabla = \nabla _{\rm rad} =
\left( d \ln T / d \ln P \right) _{\rm rad} $ and in convective regions
we use $ \nabla = \left( d \ln T / d \ln P \right) _{\rm ad}$. The
adiabatic derivative is calculated for each mass point during the
evaluation of the equation of state.

We solve these coupled equations with a variant of the relaxation
method described definitively by Kippenhahn, Weigert, \&
Hofmeister (1967) and still in extensive use today. This method
transforms the main {\it differential} equations into {\it
difference} equations by expanding them in a Taylor series about
small corrections to the initial guesses for the primary
variables. The corrections are derived iteratively by a routine
adapted from Press et al. (1987; Chapter 16.3).

\subsection{Physical Ingredients of the Models}

\subsubsection{Equation of State}

The equation of state (EOS) is a numerical function that relates the
primary variables pressure $P$, temperature $T$, and the secondary
variable density $\rho$. The EOS is that of an ideal monatomic gas with
radiation pressure. Partial ionization of the metallic species is
included but is generally unimportant at $T
> 10^7$ K in the stellar core.

\subsubsection{Opacity}

Our code computes the Rosseland mean opacity $\kappa =
\kappa(\rho,T,X,Z)$ from the OPAL tables produced at the Lawrence
Livermore National Laboratory (Iglesias \& Rogers 1992). The code calls
a subroutine provided publicly by the OPAL project to interpolate in
the tables and derive $\kappa$ for a specific combination of parameters
$(\rho,T,X,Z)$. The tables are provided at fixed grid points in T and
$R$ ($R = \rho/(T/10^6 K)^3$, not to be confused with the radius) and
there are multiple tables corresponding to different metallicities.

\subsubsection{Energy Generation}

Energy generation rates ($\epsilon$, in erg s$^{-1}$ g$^{-1}$) are
computed by a custom nuclear reaction network. The network stores
the relative populations of the relevant nuclear species and
derives from these and the reaction rates the energy generation
rates from proton-proton, CNO, and He burning at supplied values
of the primary and secondary variables. The 24 isotopic abundances
appearing in Table 1 are followed individually in a fully
implicit, time-dependent network modeled after the formalism in
Hix \& Thielemann (1999). Table~1 also lists the included
reactions, for which the reaction rates are drawn from the
canonical Caughlan \&
Fowler (1988) compilation.

The nuclear reaction network is coupled to the stellar structure
equations through the secondary variable $\epsilon$ and the abundances
of the chemical species. These are stored in a separate array and the
hydrogen, helium, and metal mass fractions, $X$, $Y$, and $Z$,
respectively, are computed from this array, with scaling to the total
density for use in the EOS and opacity routines.

\subsubsection{Convection and Mixing}

Where the temperature gradient is too steep for radiative energy
transport, convective motions must carry the energy produced in the
stellar core. We assume that the isotopic abundances are fully mixed in
convective regions and that convective regions overshoot their formally
defined boundaries. We follow the common practice (Maeder \& Meynet
1987; Schaller et al.~1992) that assumes an overshooting distance of
0.20$H_p$, where $H_p$ is the pressure scale height evaluated at the
convective boundary:
\begin{equation}
H_p = P \left| \frac{dR}{dP} \right| = \frac{P}{(GM_r/r^2)\rho}.
\end{equation}
Different assumptions for the overshooting distance may contribute
modestly to the systematic uncertainty in the behavior of the stars in
the HR diagram and their emergent spectra.

\subsubsection{Boundary Conditions}

Four boundary conditions are required to complete the system of
four equations (Eq. 1 -- 4). The outer boundary conditions
constrain the temperature and pressure at the photosphere of the
star. We assume that the temperature at the photosphere is the
effective temperature given by the relation $L = 4\pi R_*^2
\sigma_{\rm SB} T_{\rm eff}^4$.

To calculate the atmospheric pressure, we call a function that
derives $P_{\rm atm} = P(T, \rho, X, Y, Z)$ for the stellar
photosphere. Near the stellar surface, we neglect changes in mass,
radius, and luminosity and assume that $M_r = M_{*}$, $R = R_{*}$,
and $L_r = L_{*}$. This region obeys a simple relationship common
to all stratified atmospheres: $d\tau = -\kappa \rho dr
\label{eq:atmos1}$. Using this relation and combining equations 1
and 3, we derive a differential equation relating optical depth
$\tau$ and pressure $P$:
\begin{equation}
\frac{dP}{d\tau} = \frac{GM_*}{\kappa R_*^2}. \label{eq:atmos2}
\end{equation}
Then, using the Eddington approximation to radiative transport, we
write the relation:
\begin{equation}
T^4 = \frac{3}{4} \left( \frac{L_*}{4\pi \sigma_{\rm SB} R_*^2}
\right) \left(\tau + \frac{2}{3} \right). \label{eq:eddapprx}
\end{equation}

The boundary condition for the differential equation (8) is that
the gas pressure vanishes at $\tau = 0$, leaving $P = P_{\rm rad}
= \frac{1}{3}(4\sigma_{\rm SB}/c)T^{4}$. At the photosphere, $T =
T_{\rm eff}$ and solving for $\tau$ in
equation~(\ref{eq:eddapprx}) gives $\tau = 2/3$. We then integrate
equation~(\ref{eq:atmos2}) from $\tau = 0$ to $\tau = 2/3$, and
find the value of $P$ corresponding to $T_{\rm eff}$ at the
photosphere. This calculation is done at each iteration in the
approach to convergence.

At the inner boundary we use the so-called ``central expansions''
(Kippenhahn et al.~1967), which fix the luminosity and radius at the
inner mass zone.


\subsubsection{Initial Conditions}

The starting model for an evolutionary sequence is calculated at the
zero-age main sequence (ZAMS). A mass and chemical composition are
specified, and a static structure model is calculated at $t=0$. In this
case, the starting models were the ZAMS models presented by TS. We use
the initial composition $X$ = 0.76, $Y$ = 0.24 to represent the
primordial composition. For illustration, we note that a decrease in
$X$ leads to systematically higher core temperatures for stars powered
by pp and CNO burning ($\epsilon_{\rm pp} \propto X^2$ and
$\epsilon_{\rm CNO} \propto XZ_{\rm C}$).

For $M \gtrsim 15$ \Msun, ZAMS Pop III stars are not strictly
metal-free except perhaps in the early stages of their formation (El
Eid et al. 1983; Castellani et al. 1983; Marigo et al. 2000). The
stellar cores are hot enough to accumulate a small abundance of
$^{12}$C via the triple-$\alpha$ process before and during
main-sequence H-burning. Thus, the idea of a high-mass, metal-free,
main-sequence star powered solely by pp burning is misleading. In fact,
these stars derive most of their energy from CNO-burning of H (see
\S~3.2 below). We must carefully follow the production of $^{12}$C in
the core to derive the initial conditions for a main-sequence (MS) Pop
III model. In contrast, TS used {\it ad hoc} assumptions for the pre-MS
synthesis of carbon, based on the earlier calculations of El Eid,
Fricke, \& Ober (1983).

We start with the strictly metal-free ZAMS model from TS and let
it produce $^{12}$C in the core until CNO burning dominates the
energy production. Starting from the $Z = 0$ model, the star will
evolve to slightly lower luminosity and $T_{\rm eff}$ until energy
production from CNO burning exceeds that of the pp chains. At this
point, the star begins to evolve to higher luminosity but still
declines in $T_{\rm eff}$. The $^{12}$C mass fraction at this
transition is $Z_{C} \sim 10^{-12}$ (where $Z_{\odot} \sim 0.02$)
and increases slightly with mass. We take the first point where
$dL/dt > 0$ to be the starting point of the evolutionary tracks.
For $M \lesssim 15$ \Msun, luminosity increases immediately,
starting at $Z=0$.

\section{Results for Stellar Evolution at Zero Metallicity}

\subsection{Evolutionary Tracks}

\begin{figure}
\plotone{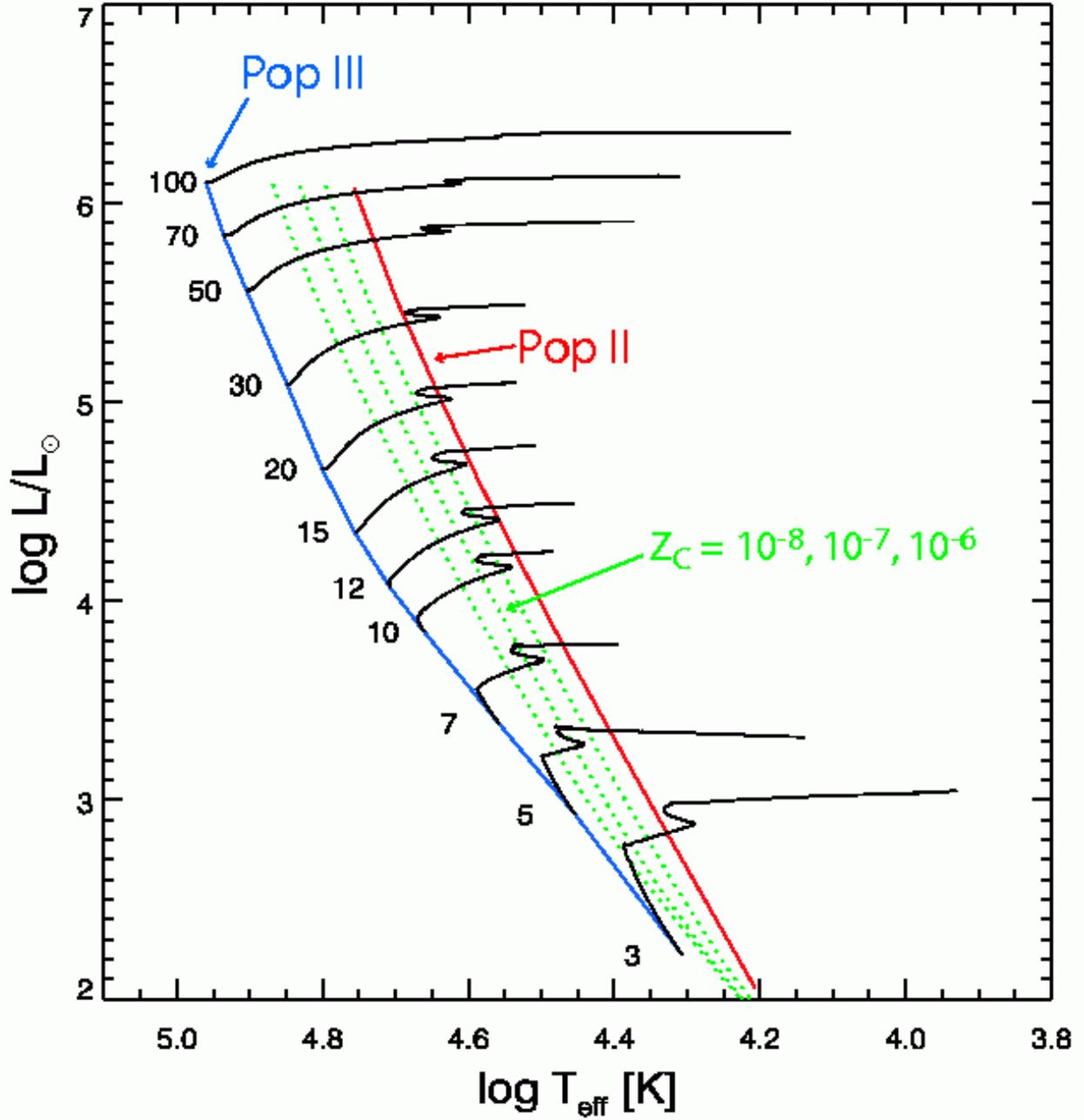}
\figcaption[The $Z=0$ evolutionary tracks in the HR diagram.]{The
$Z = 0$ stellar evolution tracks from 3 - 100 \Msun. The 1 \Msun\
track has been omitted for clarity. These tracks terminate when
the convective core expands to reach the H-rich region during core
He burning. To illustrate the stark differences between Pop III
and their metal-enriched counterparts, we mark in red the ZAMS for
Pop II ($Z = 0.001$) from Schaller et al. (1992). The Pop III ZAMS
is marked in blue, and the ZAMS for $Z_C =$ 10$^{-8}$, 10$^{-7}$,
and 10$^{-6}$ are plotted with dashed lines, with metallicity
increasing to the right.\label{fig:tracks}} \end{figure}

We have calculated evolving stellar models for $M = 1 - 100$ \Msun\
from the ZAMS past the onset of helium burning. These tracks appear in
Figure~\ref{fig:tracks}, which shows the tracks compared with a similar
Pop II ($Z = 0.001$) zero-age main sequence for $M$ = 1 -- 100 \Msun\
from Schaller et al. (1992). The large gain in effective temperature at
$Z = 0$ is readily apparent in this figure.

We have done a simple comparison of our tracks to the similar work of
Marigo et al. (2000) and Siess et al. (2002). Our tracks match theirs
on the H-burning main-sequence, and for He-burning at $M > 10$ \Msun.
We find effective temperatures that are 3 - 5\% lower and H-burning
lifetimes that are 1 - 12 \% shorter (Table 2). These differences
illustrate that individual choices for parameters, numerical
techniques, and the specific implementation of physical processes
(e.g., convection) or boundary conditions can create small differences
in the final results. Our models diverge from Marigo et al. (2000) in
the details of evolution in the HR diagram during the He-burning phases
for $M \leq 10$ \Msun, perhaps owing to differences in our treatment of
convection. Since we are concerned with the spectral evolution for the
first 10 Myr, which is dominated by the H-burning MS of massive stars,
we have not attempted to resolve these differences here. Simple testing
indicates that the time evolution of ionizing photons and broadband
spectra are similar between the two sets of tracks, except for a
roughly 30\% smaller He II ionizing photon production from our models,
due to the slightly cooler effective temperatures and the sensitivity
of 4 Ryd radiation to $T_{\rm eff}$.  We note also the studies of Pop
III stars over a broad mass range by Schaerer (2002), and Bromm et al.
(2001), who published static models of zero-age Pop III stars from 100
- 1000 \Msun.

Our tracks do not complete core He burning because of a peculiar
evolutionary feature noted also by Marigo et al. (2000). During
core He burning, the core begins to expand and at some point
reaches the boundary of the H-depleted region. At this point, the
core H abundance rises from 0 to $\sim 10^{-4}$ from one discrete
timestep to the next, owing to the assumption of instantaneous
complete mixing in the convective core. This assumption is clearly
inadequate here, since the added H is probably burned quickly at
the edge of the convective region in a continuously evolving star.
An {\it ad hoc} treatment of this phenomenon that forbids
expansion of the convective core past this boundary is unlikely to
represent accurately the true behavior of the stellar interior in
this stage. Since this evolutionary phase represents a small
fraction of He burning, which is itself $\lesssim$10\% of the
stellar lifetime, we terminate our tracks when the boundary of the
convective core reaches the H-rich region.

Figure~\ref{fig:tracks} shows that, for $M \gtrsim 12$ \Msun, the
stars evolve uniformly towards lower $T_{\rm eff}$ and higher
luminosity. These stars are powered by core CNO burning, catalyzed
by the small C fraction built up there by 3$\alpha$ burning. For
$M \lesssim 12$ \Msun, the stars start on the ZAMS with $Z=0$ and
first evolve to higher $T_{\rm eff}$ via pp burning while the core
C abundance rises. When the $^{12}$C mass fraction reaches $Z_C
\sim 10^{-9}$, the star is powered primarily by CNO burning and
begins to evolve to lower $T_{\rm eff}$.

To assess the differences in ionizing photon emission and some
observational characteristics between Pop III and Pop II, we have
constructed zero-age main sequences for initial carbon abundances
of $Z_C = $ 10$^{-8}$, 10$^{-7}$, and 10$^{-6}$. These appear in
Figure~\ref{fig:tracks}. These unevolved populations may represent
the intermediate stellar generations between Pop III and Pop II.
We construct model spectra for these populations as for the Pop
III evolving cluster (see \S~4).

We have neglected mass loss in calculating our stellar evolution models
and spectra. Mass loss is believed to be driven in massive stars by
radiation pressure on blanketed metal lines in the outer reaches of the
stellar atmospheres. In the metal-free case, this pressure is likely to
be minimal (Kudritzki 2000), and the constant-mass assumption is a good
approximation. Marigo et al. (2000) used a simple test of Eddington
luminosity to argue that massive Pop III stars may experience modest
mass loss late in their He-burning phases. Even if this occurs, it is
unlikely to affect the composite spectra we produce here and can be
safely ignored.  The uncertain role of mass loss can be viewed as
one of the major systematic uncertainties remaining in the study of
metal-free stars. If it is strong, main-sequence mass loss drives the
star to hotter $T_{\rm eff}$ on the main sequence - these effects can
be seen in the tracks of El Eid et al. (1983) and Klapp (1983).

\subsection{Why Pop III Stars are Hot}

\begin{figure}
\plotone{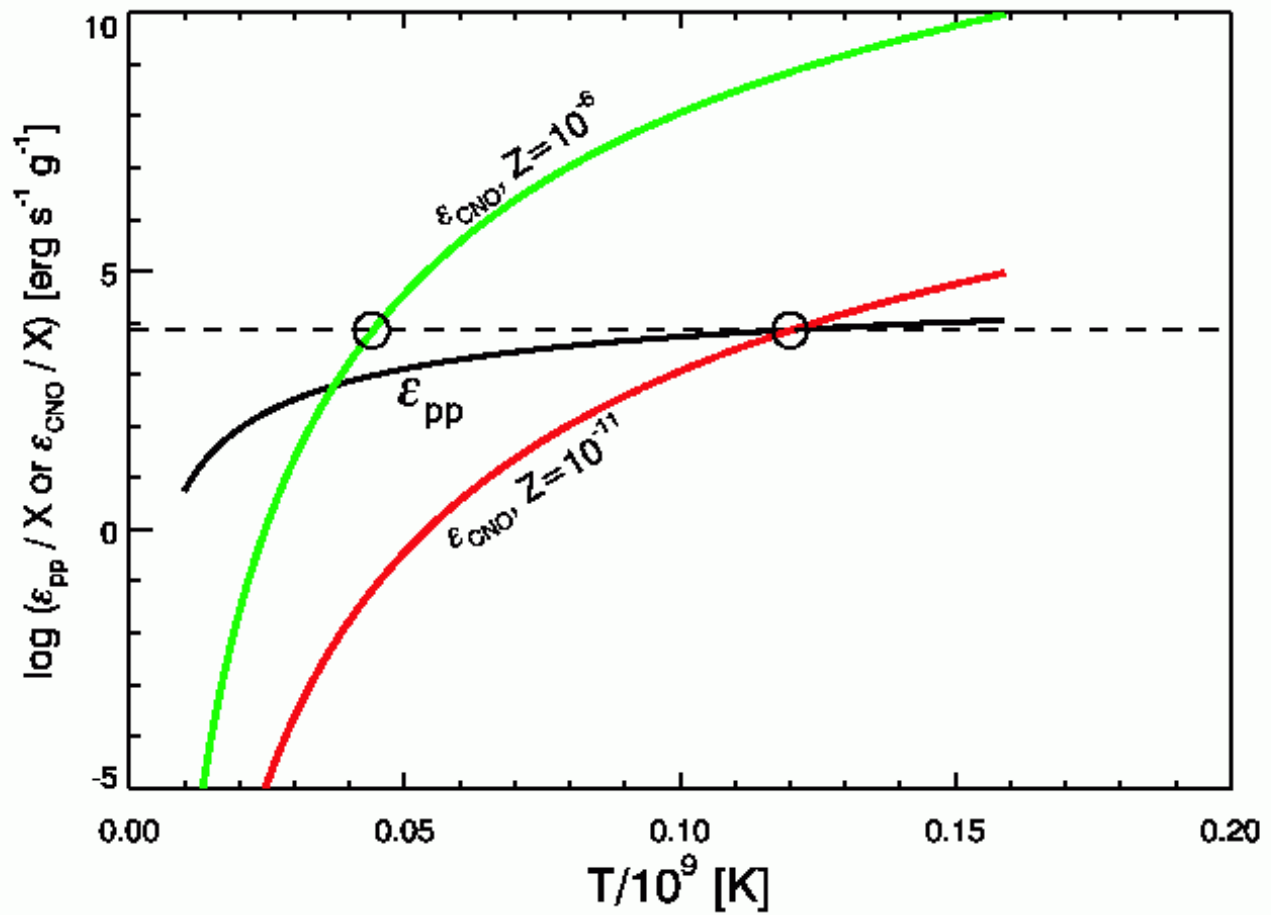}
\figcaption[A comparison of pp to CNO energy
generation.]{Energy generation rates ($\epsilon$) illustrate why
metal-free stars must attain unusually high temperatures to
support their mass against gravity. The black curve shows the pp
rate for a typical H mass fraction $X$ = 0.7. The green and red
curves show the CNO rate for a fiducial low metallicity and for
Pop III, respectively. The horizontal dashed line marks a typical
(but arbitrary) rate for the interior of a massive ($M
> 20$ \Msun) star. The circles mark the points where the
respective burning processes produce the desired total $\epsilon$.
See text for discussion. \label{fig:epscompare}}\vspace{0.05in}
\end{figure}

The distinctive spectral appearance of Pop III stars is due directly to
their unusually hot surfaces (TS). These stars are roughly two times
hotter and five times smaller than their Pop II counterparts of the
same mass. There are two main reasons for these differences, both
related directly to the chemical composition. First, nuclear burning in
a metal-poor isotopic mix must occur at higher temperatures to derive
the same energy per unit mass as at higher metallicity. The change in
nuclear energy production rates ($\epsilon$) is illustrated in
Figure~\ref{fig:epscompare}, with some characteristic parameters. The
highest curve (green) is $\epsilon_{\rm CNO}$ for a metallicity of $Z_C
= 10^{-6}$. At this metallicity, CNO burning at a typical core
temperature of $4 \times 10^7$ K can achieve the typical $\epsilon$ for
a particular location in the stellar interior. If this star is
restricted to pp burning (black curve) or to CNO burning at a very low
C abundance (red curve), it must do so at temperatures above $10^8$ K.
These latter two cases illustrate Pop III nuclear burning. This effect
accounts for most of the large changes in stellar structure as
metallicity approaches $Z = 0$. Second, and less important to the
structure of Pop III stars, the opacity of stellar material is reduced
at low metallicity, permitting steeper temperature gradients and more
compact configurations at the same mass (see Figure 3).

\begin{figure}
\plotone{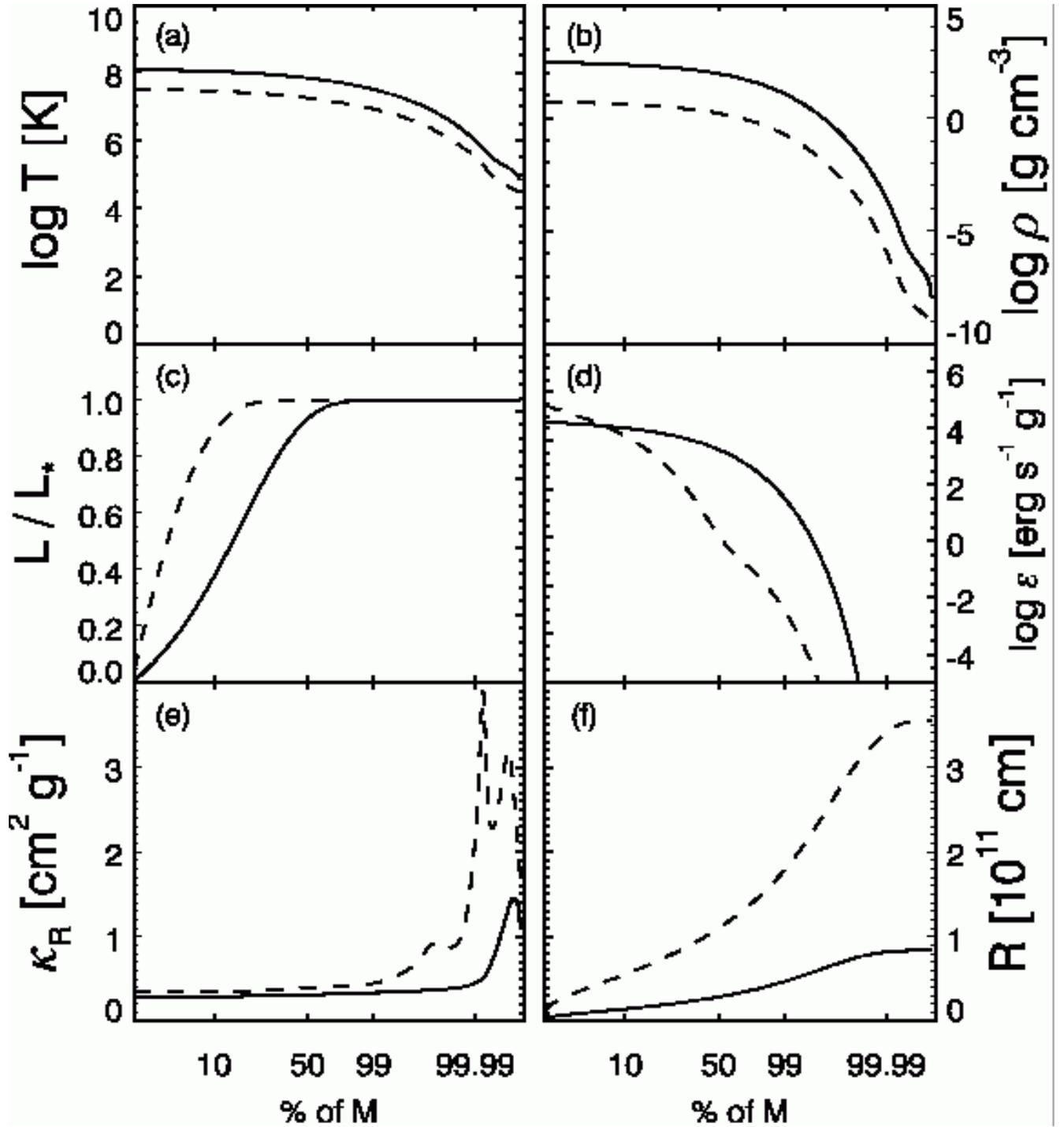}
\figcaption[Comparison of stellar interiors for Pop III and Pop II.]{A
comparison of the interior structure of a Pop III star (solid curves)
with a Pop II star (dashed curves), both of $M_* = 15$ \Msun. The Pop
III star is smaller, hotter, and denser. This comparison demonstrates
that the gain in $T_{\rm eff}$ arises not from changes in the opacity
with metallicity, which differs substantially only in the outer 1\% of
the total mass, but rather from changes in the available fuel source in
the stellar core. See \S~3.2 for discussion.
\label{fig:varcompare}}\vspace{0.05in}
\end{figure}

The effects of reduced opacity can be seen in
Figure~\ref{fig:varcompare}, which shows the run of six important
physical quantities in Pop II (dashed) and Pop III stars (solid) of 15
\Msun. Pop III stars are more centrally condensed, with hotter, denser
cores, and hotter surfaces. Throughout 99\% of the stellar material,
the opacity (lower left) has a weak dependence on metallicity because
it is dominated by scattering off electrons donated primarily by H. In
the outer envelope, which comprises 1\% of the stellar mass, the
opacity is provided by the bound-free and bound-bound transitions of
H~I, He~I, and He~II. Here the opacity differs substantially between
Pop II and Pop III, but it has little effect on the overall stellar
structure.

\section{Model Atmospheres and Synthetic Spectra}

\subsection{Model Atmospheres}

\begin{figure}
\plotone{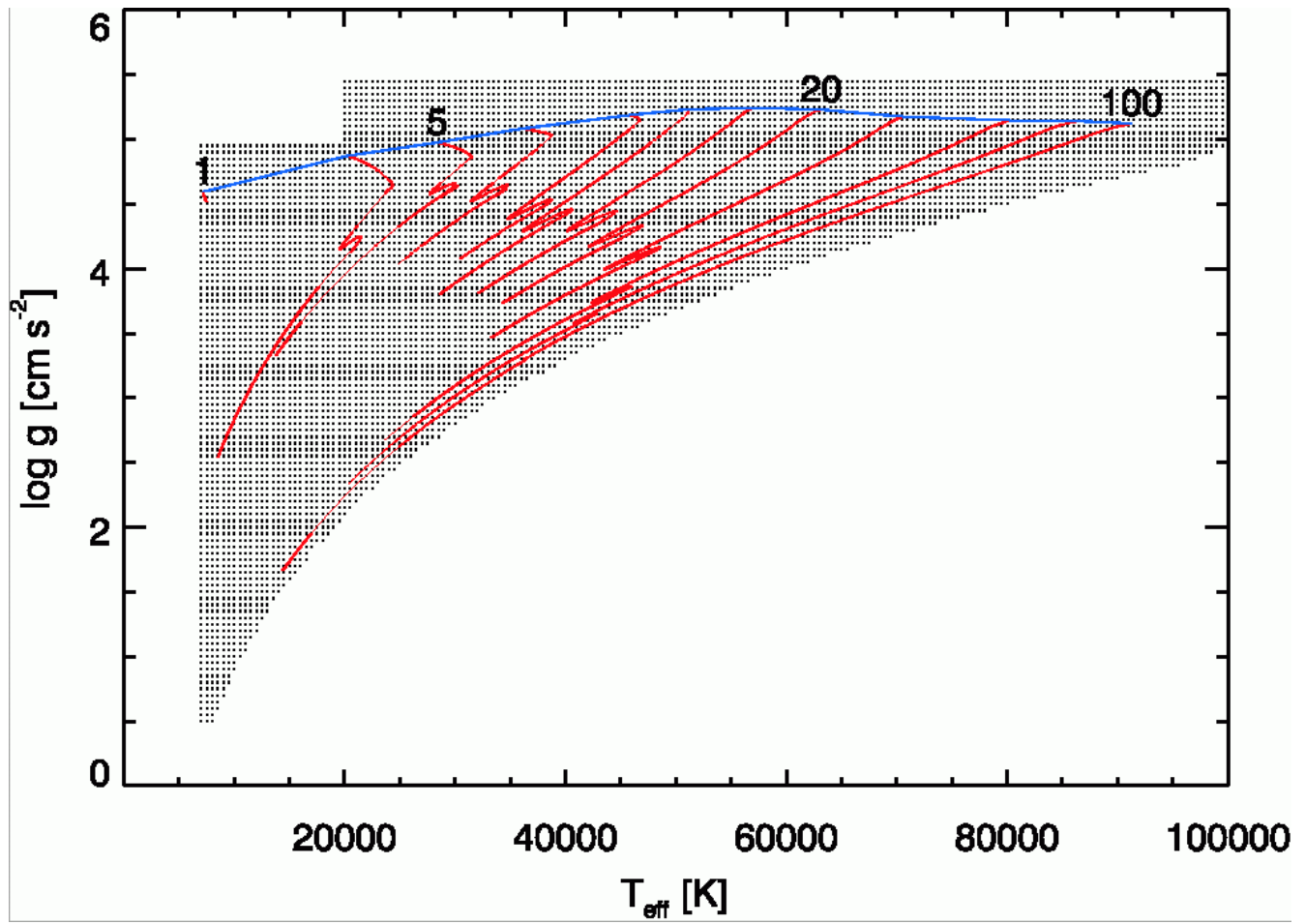}
\figcaption[The model atmosphere grid.]{Parameter grid ($T_{\rm eff}$,
$\log g$) of the TLUSTY NLTE model atmosphere grid with the Pop III
tracks overlaid. The grid boundaries were chosen specially to cover the
tracks with a minimum of extraneous points. The grid spacing is $\Delta
T_{\rm eff} = 500$ K and $\Delta (\log g) = 0.05$. Only the first
$10^9$ yr of the 1 \Msun\ track is accommodated by the grid. The Pop
III ZAMS is marked in blue.  \label{fig:gridfig}}
\end{figure}

The model atmosphere grid was calculated using the publicly
available TLUSTY code (Hubeny \& Lanz 1995). These models are
parameterized by the effective temperature at the stellar
photosphere ($T_{\rm eff}$) and the surface gravity ($g$). Model
spectra at the grid points are calculated for $\lambda = 1 -
10,000$ \AA\ by the companion program SYNPLOT. For high accuracy
and coverage we chose $\Delta T_{\rm eff} = 500$ K and $\Delta
(\log g) = 0.05$. Figure~\ref{fig:gridfig} shows the range of the
model grid with the evolutionary tracks overlaid. At low
temperatures ($T_{\rm eff} < 10,000$ K) the model atmospheres are
allowed to have small convective regions to promote convergence.
This modification has little effect on their emergent spectra.
Some stars develop convective envelopes during late H burning, but
these regions lie far from the nuclear burning core and
thus always maintain their metal-free composition. 

\subsection{Population Synthesis}

To produce model spectra and calculate the cosmological effects of
Pop III stars we must create synthetic clusters from the discrete
set of tracks presented above. We desire smoothly varying
quantities such as ionizing photon flux and broadband colors, so
we must convert the tracks at discrete masses into more continuous
variables.

Constructing a complete synthetic population from evolutionary
tracks requires careful calculation of the tracks for masses that
are not explicitly computed. For example, if one has tracks at
100, 80, and 60 \Msun, extrapolating the behavior of the 90 \Msun\
track from 60 and 80 \Msun\ leads to large errors, and
interpolation of the 100 and 80 \Msun\ tracks is not accurate
after the last timestep on the more massive track. We cannot
simply use the explicitly calculated tracks to represent broad
mass bins ($\Delta M \sim$20 \Msun), because this leads to sharp
discontinuities in the calculated spectra and derived quantities.
We use small mass bins of 1~\Msun, which are preferred for
adequate accuracy and smooth behavior of the results. To resolve
these issues we use a ``lifetime-proportional'' interpolation
scheme to create synthetic evolutionary tracks for the
intermediate masses between the explicitly calculated tracks.

The evolutionary tracks at the explicitly calculated masses are
followed beyond the onset of helium burning, a phenomenon that occurs
at all masses and occupies $\lesssim 10$\% of the stellar lifetime.
Since later stages (C-burning, Si burning) are even shorter and
contribute little to the spectral evolution of a cluster, we assume
that He burning is the last evolutionary phase and that its end
terminates the star's life. We also rely on the fact that the changes
in shape and time-dependence of the tracks vary smoothly with mass. We
can use these important features to construct interpolated tracks at
intermediate masses.

Taking the run of $T_{\rm eff}(t)$ and $L(t)$ to be a track, we first
remap the explicit tracks to a new time coordinate - the fraction of
the stellar lifetime achieved to that point, ranging from 0 to 1. Then,
a vector of lifetimes for each mass bin is interpolated from the
lifetimes of explicit tracks. Finally, we interpolate intermediate
tracks between the explicit tracks at the same fraction of their
lifetimes. For instance, to find ($L$, $T_{\rm eff}$) for a 90 \Msun\
star at 70\% of its interpolated lifetime, we interpolate in mass
between the 80 \Msun\ and 100 \Msun\ explicit tracks at 70\% of their
lifetimes. This scheme minimizes interpolation errors and generates
smoothly varying model spectra and derived quantities.

Using the interpolated tracks and model atmospheres, we construct
synthetic stellar populations by specifying a stellar initial mass
function (IMF) and adding stellar spectra together. Because we lack
detailed knowledge of the IMF for $Z=0$ star formation, we assume that
the IMF is a power law, $dN/dM \propto M^{-\alpha}$, with the Salpeter
slope, $\alpha = 2.35$. To populate the IMF, we construct 99 mass bins
and populate them with the number of stars of that mass, normalized to
a total mass of 10$^6$ \Msun, which is standard practice. The mass bins
are 1 \Msun\ wide and range from 1 - 100 \Msun. Model spectra from the
atmosphere grid are assigned to the bins and summed over all masses.
The product of this process is a composite spectrum for the whole
cluster. Schaerer (2002) and Bromm et al. (2001) performed similar
calculations for top-heavy IMFs ranging up to 1000 \Msun.

We adopt a model of Pop II starburst clusters as a reference spectrum
for comparison with our evolving Pop III models. In the present
paper, this reference model represents a baseline spectrum for
comparison to the $Z = 0$ models. In Paper II, we use this spectrum as
a proxy for the ``second generation'' of stars incorporating the metals
produced by the first. For the reference population, we use a model
with $Z = 0.001$, $M = 10^6$ \Msun\ in stars, and a Salpeter IMF, that
was custom-generated by the Starburst99 population synthesis code
(version 3.1, Leitherer et al. 1999). Because this suite of models is
widely used in the community, we compare all our model spectra and
observational predictions to a Pop II model generated by their code. We
note further that we do not include Wolf-Rayet (WR) stars in these
comparison models to make our comparisons as direct as possible. WR
stars are known to produce large amounts of He II ionizing photons in
the local universe (Schaerer \& Vacca 1998), but their at high redshift
($z > 5$) and low metallicity is regarded as unlikely since the WR
phenomenon requires metal-line opacity to drive strong stellar mass
loss. Recent studies of stellar mass loss at low metallicity
(Kudritzki 2002) have shown that its importance is minimized at
extremely low metallicity, with a factor of 10 drop in mass-loss rate
from $Z = 10^{-3}$ to 10$^{-4}$. Nonetheless, we include in our
comparisons below some results for Pop II with WR stars. 

\begin{figure}
\centerline{\epsfxsize=0.55\hsize{\epsfbox{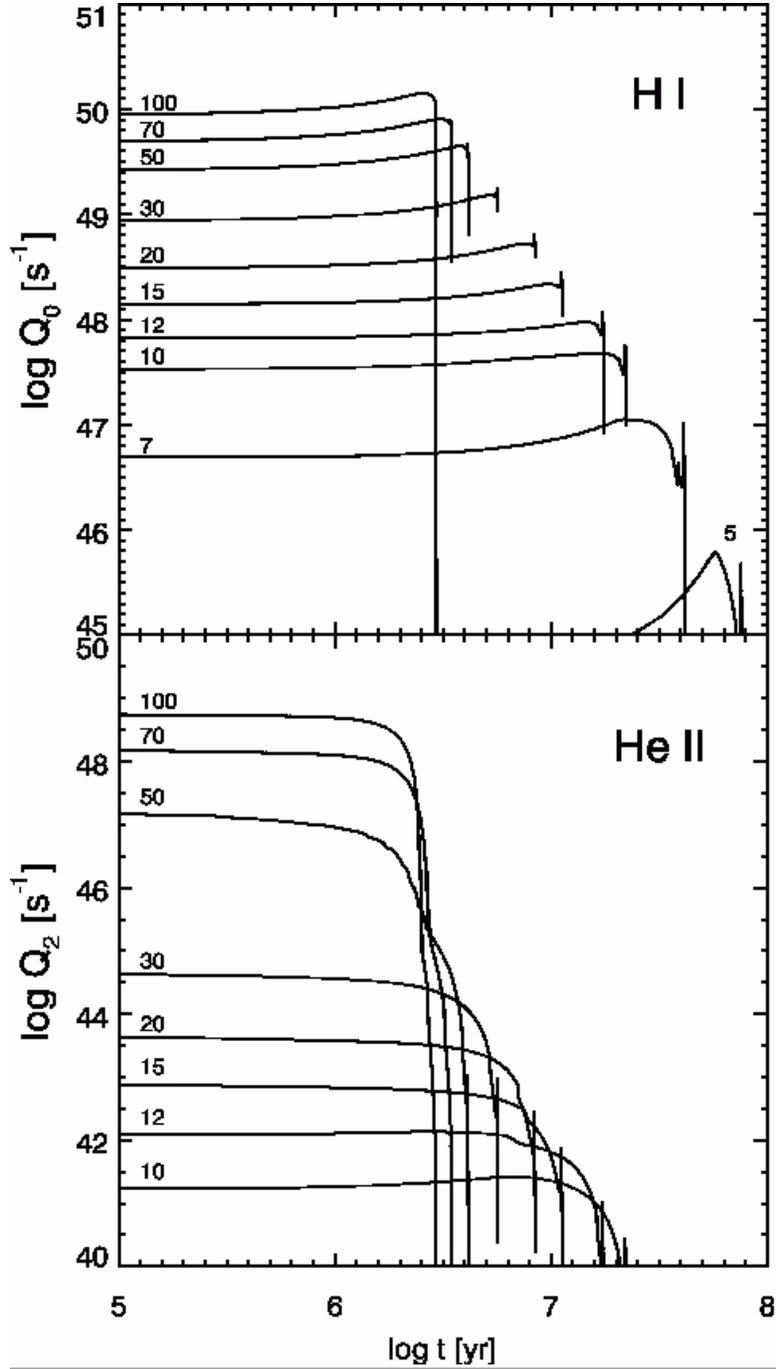}}}
\figcaption[Ionizing photon production from Pop III stars.]{Ionizing
photon production rates $Q_0$ (1 Ryd; upper panel) and $Q_2$ (4 Ryd;
lower panel) for individual stars from the tracks in
Figure~\ref{fig:tracks}.  The tracks are labeled with their stellar
mass in \Msun.  High-mass, zero-metallicity stars are strong sources of
H~I and He~II ionization. In particular, they emit $10^5$ times more
radiation in the He~II continuum than Pop II stars.
\label{fig:qfig}}
\end{figure}

For modeling cosmological reionization, radiative feedback, and
the observational signatures of metal-free stars, we make use of
the important quantities $Q_i$, which express the ionizing photon
output per second in the H~I, He I, and He~II continua ($i$~=~0,
1, 2, and $h\nu_i$ = 13.6, 24.6, 54.4 eV, respectively):
\begin{equation}
Q_i = 4\pi R_{*}^2 \int^{\infty}_{\nu_i} \frac{F_\nu}{h\nu} d\nu,
\end{equation} where $F_{\nu}$ is the spectrum of the star in units
of erg cm$^{-2}$ s$^{-1}$ Hz$^{-1}$. The time-dependent values of
$Q_0$ and $Q_2$ are displayed for the individual stars in
Figure~\ref{fig:qfig} to show the behavior with mass.

Once the individual $Q_i$ have been weighted by the IMF, we arrive at
the cluster $Q_i$ seen in Figure~6. There are several features of this
figure that deserve comment. First, note the starting (ZAMS) values of
$Q_0$ and $Q_2$ for the Pop III and comparison Pop II clusters. The Pop
III cluster has $Q_0$ that is 60\% larger than the Pop II cluster.
Second, the ionization of the Pop III cluster persists at a high level
far longer than the Pop II cluster, owing to the strong gain in
ionization for $M = 10 - 25$ \Msun\ stars. The Pop III cluster takes 10
Myr to decline to 1/10 its ZAMS value in $Q_0$, twice as long as Pop
II. Finally, note the strong ionization above the He~II edge at 4 Ryd.
At the Pop III ZAMS, we find that $Q_2$/$Q_0$ = 0.02, with a lifetime
of 2.5 Myr, characteristic of $M \geq 40 - 50$ \Msun\ stars (see Figure
5). This form of the output from our models can be placed directly into
cosmological calculations of IGM reionization that incorporate
realistic models of evolving spectra from Pop III stars (see Paper II).

For the ZAMS spectra at $Z_C = $ 10$^{-8}$, 10$^{-7}$, and 10$^{-6}$,
we find $Q_0 = (8.9,8.4,7.9) \times 10^{52}$ s$^{-1}$, respectively,
intermediate between the Pop II and Pop III values. There is a sharp
decrease in their $Q_2$, however, with $Q_2 = 8.4 \times 10^{49}$, $2.1
\times 10^{48}$, and $4.8 \times 10^{47}$ s$^{-1}$, respectively. These
quantities are compared with the evolving $Q_2$ for Pop III and Pop II
in Figure~6.

In Paper II, we examine the effects of Pop III clusters on the
reionization of the IGM. These tests use a quantity $\eta$, which
expresses the ratio of He~II to H~I column densities in an ionized IGM
(Fardal, Giroux, \& Shull 1998). This ratio can be predicted for a
specific distribution of IGM clouds, given an input source spectrum,
which is assigned an intrinsic $\eta$ (prior to IGM filtering) based on
the shape of its spectrum above 1 Ryd. A composite QSO (Zheng et al.
1997; Telfer et al. 2002) with a power law spectrum, $F_{\nu} \sim
\nu^{-\alpha}$, and $\alpha = 1.8$ has an intrinsic $\eta = 20$. At the
ZAMS, the Pop III cluster has $\eta = 10$, harder than the mean QSO and
similar to a power-law spectrum with $\alpha = 1.3$. For the
low-metallicity ZAMS at $ Z = 10^{-8}, 10^{-7}$, and 10$^{-6}$, we find
intrinsic $\eta$ = 33, 80, and 180, respectively. We make use of this
ratio in Paper II to explore the fate of IGM regions ionized by Pop III
stars.

\section{Observational Signatures}

\subsection{General Approach}

Observational searches for high-redshift galaxies work best when
equipped with diagnostic tools that provide detailed relationships
between observable signatures and underlying physical phenomena.
To this end, we use our models of Pop III clusters to derive
predictions for the observed spectral energy distributions (SED)
of the first stars. We consider two classes of observation; the
first seeks the nebular emission line signatures of Pop III stars,
and the second exploits their broadband SED and colors. We take up
the nebular emission first because its presence affects the
broadband signatures as well.

\subsection{Emission-line Diagnostics}

Emission-line signatures are proving to be the primary means of
finding galaxies at redshifts beyond $z = 5$. Deep, blind searches
for Ly$\alpha$ emitters have turned up large populations of
galaxies at these epochs. These surveys use either blind
multi-slit spectrographic exposures to detect single emission
lines over a broad redshift range (Hu et al. 1999), or they take
narrowband images to probe deeply at selected redshifts (Rhoads et
al. 2000).

TGS derived a relationship for the observed He~II emission-line flux
from metal-free stars as a function of star formation rate (SFR). With
the evolved spectra of metal-free stars we can update these earlier
calculations. We assume that few He~II ionizing photons escape the
galaxies where they are produced ($f_{\rm esc}$ = 0.025; see Paper II)
and that the sources themselves contain no dust. We assume Case B
recombination at 20,000 K, corresponding to a higher nebular
temperature in low-metallicity gas. As in TGS, we adopt
$j_{4686}/j_{\rm H\alpha}$ = 0.66, $j_{1640}/j_{\rm H\alpha}$ = 4.6,
and $j_{3203}/j_{\rm H\alpha}$ = 0.28 (Seaton 1978). TGS used an {\it
ad hoc} scale factor, $f_{\rm evol}$, to account for stellar evolution
effects, which reduce the He~II ionizing flux over time as high-mass
stars evolve off the main sequence. We need not make this approximation
here, since we have directly calculated evolving spectra of metal-free
stars. For both continuous and instantaneous star formation, we assume
that $L_{\rm H\alpha} = 1.27 \times 10^{-12} Q_0$ erg s$^{-1}$ (i.e.,
there are 0.424 H$\alpha$ photons emitted for each H-ionizing photon).
Similar coefficients are derived for the other H~I lines from Case B
recombination theory (Osterbrock 1989). Note that Leitherer et al.
(1999) use a coefficient of $1.36 \times 10^{-12}$, corresponding to
nebular gas at 10$^4$ K. We then multiply the H$\alpha$ luminosity by
the emissivity ratio $j_i/j_{H\alpha}$ to derive $L_{i}$ in a He~II
line. In Figure~\ref{fig:obs} we plot the predicted line fluxes for
Ly$\alpha$ and He~II $\lambda$1640 as a function of redshift.

For clusters with significant ionizing photon production, we must also
account for continuous nebular emission from the H~I two-photon process
and direct recombinations to upper levels of H~I, He~I, and He~II. We
construct a wavelength-dependent emission spectrum, in units erg
s$^{-1}$ \AA$^{-1}$:
\begin{equation}
F_{\lambda}^{\rm cont} = \frac{c\gamma_i}{\lambda^2 \alpha^{B}_{H}}
(1-f_{\rm esc }) Q_0,
\end{equation} where the term $\gamma_i$, in units erg cm$^{3}$ s$^{-1}$ Hz$^{-1}$,
incorporates contributions from H~I, He~I, He~II, and the H~I
two-photon continuum and $\alpha_{H}^{(B)}$ is the Case B recombination
rate coefficient for hydrogen. The emission coefficients for an
electron temperature of 20,000 K are taken from Aller (1984, Table
4.9). This additional emission has a large effect on the broadband
spectral energy distribution of the synthetic clusters, shifting energy
from the Lyman continuum to the rest-frame optical bands. This effect
is included in the broadband color and equivalent width predictions
presented below.

In general, we can parameterize the escape of radiation with the
quantity $f_{\rm esc}$, the fraction of ionizing photons that escape
from the host halo into the IGM. To be consistent with the typical
reionization models presented in Paper II, we uniformly assume $f_{\rm
esc} = 0.05$ for 1 -- 4 Ryd and $f_{\rm esc} = 0.025$ for $h\nu > 4$
Ryd (Dove, Shull, \& Ferrara 2000; Tumlinson et al. 1999), leaving most
ionizing photons at all wavelengths to excite observable nebular
emission. We return to the issue of the escape fraction in Paper II,
where we discuss justifications for different escape fractions at 1 and
4 Ryd.

In Figure~\ref{fig:obs} we show the flux of the $\lambda 1640$ and
Ly$\alpha$ lines from the Pop III cluster. The green curves represent
continuous star formation at 1 and 40 \Msun\ yr$^{-1}$. The higher rate
corresponds to the star formation rate (SFR) inferred by Hu et
al.~(2002) for a Ly$\alpha$ emitter discovered at $z=6.56$ (Ly$\alpha$
flux marked with the green square). The red curves represent the He~II
$\lambda$1640 \AA\ flux from an instantaneous burst of $M = 10^6$
\Msun\ at $t = 0$ (solid) and 1 Myr after the burst (dashed).
Figure~\ref{fig:obs} also shows the imaging and spectroscopic
sensitivity requirements for the planned 6 m aperture {\it Next
Generation Space Telescope} (NGST). These assumed limits are from the
NGST design specifications and are 2 nJy for low-resolution (R = 5)
imaging and 100 nJy for spectroscopy (R = 1000), and both assume S/N =
10 and a 100,000 s exposure time. These comparisons show that the
planned NGST will easily discover this signature of the first stars.

In Case B recombination theory, roughly two thirds of all H
recombinations result in the emission of a Ly$\alpha$ photon (Spitzer
1978), and these are known to escape from galaxies and serve as a
strong signature of high-redshift star formation (Hu et al. 1999;
Rhoads et al. 2000). For the Pop III cluster presented here, the
Ly$\alpha$ line is intrinsically stronger than in Pop II.
Figure~\ref{fig:eqw} shows the predictions for the time-dependent
Ly$\alpha$ equivalent width, $W_{\rm Ly\alpha}$, for Pop II and Pop III
assuming Case B recombination theory and $T = 20000$ K.  These curves
are technically upper limits since the Pop II cluster may suffer dust
attenuation and the Pop III Ly$\alpha$ line may be absorbed by neutral
H in the pre-reionization IGM (cf. Loeb \& Rybicki 1999). Equivalent
width has the additional benefit of being independent of the total
cluster mass for an instantaneous burst. Thus, we confirm the
suggestion of other studies (TS, Bromm et al. 2001; Schaerer 2002) that
the Ly$\alpha$ signatures of the first stars will be unusually
strong\footnote{We note here that our Ly$\alpha$ predictions lie well
below those of Schaerer (2002) for similar conditions. We use an
average flux in the 1200 - 1250 \AA\ range to represent the continuum
flux. This method approximates the observed quantities in surveys such
as LALA. If intrinsic Ly$\alpha$ absorption is present, our prediction
is an overestimate. We can reproduce Schaerer's prediction only by
assuming the continuum flux is represented by the core of the model
Ly$\alpha$ absorption line.}. The Large Area Ly$\alpha$ Survey (LALA;
Malhotra \& Rhoads 2002) found a median $W_{\rm Ly\alpha}$ of 400 \AA\
in a sample of 157 galaxies with $z = 4.5$. More than 60\% of their
detections have $W_{\rm Ly\alpha} > 240$ \AA, the approximate maximum
attainable by Pop II clusters (Charlot \& Fall 1993).

We also plot in Figure~\ref{fig:eqw} the instantaneous equivalent
widths $W_{\rm Ly\alpha} =$ 520, 430, and 360 \AA, for the $Z_C =
10^{-8},10^{-7}, {\rm and} 10^{-6}$ ZAMS spectra, respectively. The
points for $Z_C = 10^{-7}$ and $10^{-6}$ straddle the median LALA
equivalent width, indicating that this sample of Ly$\alpha$ emitters
may have a low metallicity. These results suggest that populations of
metal-free stars, or perhaps stars of the first metal-enriched
generation, have already been found. In future ground- and space-based
searches beyond the optical bands, Ly$\alpha$ emission will probably
serve as the ``signpost'' of metal-free galaxies. This result could
then be confirmed through follow-up detection of the He~II lines, which
are weaker than Ly$\alpha$ and unique to the metal-free regime.

The sensitive dependence of the Ly$\alpha$ equivalent width on
metallicity (and IMF; see Bromm et al. 2001) suggests that the
metallicity evolution of star formation may appear in the observed
Ly$\alpha$ lines at high redshift. If there is a well-defined Pop III
epoch, characterized by a rapid transition to $Z_C = 10^{-6}$ or higher
as the second stellar generation forms from gas enriched by the first,
there could be a bimodal distribution of Ly$\alpha$ equivalent widths
with redshift. The peak may appear near the $\sim$ 1000 \AA\ prediction
here, nearer the $\sim$ 3000 \AA\ predicted for the high-mass IMF
adopted by Bromm et al. (2001), or somewhere in between. Thus,
Ly$\alpha$ itself may serve as a diagnostic of the evolution of metals
in galaxies, up to at least the redshift of H reionization, where the
damping wing of Gunn-Peterson absorption in the IGM may play an
uncertain role in obscuring (Miralda-Escud\'{e} 1998) or transmitting
(Loeb \& Rybicki 1999; Haiman 2002) the galactic Ly$\alpha$ line.

\subsection{Broadband Spectra and Colors}

Broadband filter search techniques have proven successful in finding
high-redshift galaxies. In particular, the ``Lyman-break'' technique
exploits the fact that the IGM absorbs the spectrum of a galaxy
shortward of 1216(1+$z$) \AA. This powerful technique has uncovered
thousands of galaxies at $z > 3$ (Steidel et al. 1999).

The high effective temperatures of Pop III stars make their integrated
cluster spectra intrinsically bluer than their Pop II counterparts.
However, the most distinctive feature of the Pop III spectrum is the
large gain in H~I and He~II ionizing photons, which unfortunately are
emitted in the spectral region that is heavily absorbed by the IGM. To
account for this obscuration, we adopt a stochastic model of the IGM
based on the work of Fardal, Giroux, \& Shull (1998), who used Monte
Carlo techniques to construct model sightlines containing realistic
distributions of H~I Ly$\alpha$ forest, Lyman limit, and damped
Ly$\alpha$ clouds. From a sample of 1000 sightlines we construct a mean
transmission function for three redshifts, $z$ = 4, 5, 6. This
realistic treatment of the IGM leaves approximately $\sim$ 30\% of the
emergent flux between 912(1+$z$) and 1216(1+$z$) \AA. This mean
transmission function is included in the broadband signatures derived
here.

In Figure~\ref{fig:spec+emiss}, we illustrate Pop II and Pop III
spectra at high redshift. We also plot the ranges of the standard
photometric bandpasses. Emission lines are included in these colors and
make a modest contribution. Although the Pop III spectrum is
intrinsically bluer in the accessible spectral regions, the
reprocessing of ionizing photons into nebular emission makes up for
this difference. Figure~\ref{fig:colors} illustrates this similarity
between Pop III and Pop II over a range of redshifts and cluster ages
($t$ = 0 -- 10 Myr). This substantial degeneracy in the broadband
colors may make distinguishing Pop III more challenging for current and
planned optical and IR instruments than was previously appreciated.
Confusing color signatures may leave the unusual Pop III emission-line
signatures as the only means of identifying the first stars.

\section{Conclusions}

We have presented evolving spectra of metal-free stellar
populations, based on newly calculated evolutionary tracks. We
find that the evolution of Pop III stars follows the general
patterns obeyed at higher metallicity, but with an overall shift
to higher temperature. This gain in core and surface temperatures
is primarily a result of the restricted abundance of $^{12}$C in
primordial stars. We have used a grid of non-LTE model atmospheres
to produce evolving spectra of synthetic Pop III clusters. These
models have been used to compute the broadband colors and
emission-line signatures from metal-free stellar clusters at high
redshift.

Our specific conclusions fall into two categories. First, the
unique composition of the first stars has the following important
effects on their spectra:
\begin{itemize}
\item They produce 60\% more H~I ionizing photons than their
      Pop II counterparts.
\item They can produce up to $10^5$ times more He~II ionizing photons
      than Pop II, which may lead to unusual radiative feedback effects on the
      IGM.
\end{itemize}

We wish to reemphasize here the uncertainty associated with
stellar mass loss in evaluating the latter conclusion. If the ``second
generation'' of stars has $Z \sim 0.0001 - 0.001$ and perhaps
significant populations of WR stars, the gain in He II ionizing photon
production for a continuous star formation model including WR stars is
reduced to about a factor of 10 (see Figure 6). However, since the
importance of WR stars at extremely low metallicities is in doubt, we
conclude that Pop III stars will be far more efficient than their Pop
II counterparts at ionizing He II. The consequences of our models for
reionization are explored more fully in Paper II, which uses these
model spectra and a semi-analytic reionization model to assess the
importance of the first stars for full or partial reionization of H~I
and He~II in the IGM.

The peculiar properties of metal-free stars entail unusual
observational signatures:
\begin{itemize}
\item They are expected to have unusually high equivalent
      widths of Ly$\alpha$ and the He~II recombination lines,
      the most distinctive signature of Pop III stars and their nebular emission.
      These lines will probably be the primary means of detecting and
      identifying the first stars.
\item While their intrinsic spectra are significantly bluer
    than their Pop II counterparts, their broadband colors are
    similar, owing to the reprocessing of Lyman-continuum photons
    into continuous nebular emission.
\end{itemize}

Recent Ly$\alpha$ surveys suggest that metal-free stars may have
already been found. In future ground- and space-based searches beyond
the optical bands, Ly$\alpha$ emission will probably serve as the
``signpost'' of metal-free galaxies. Follow-up detections of the He~II
lines could confirm this result. The substantial degeneracy in the
broadband colors will probably leave emission line techniques as the
best way of distinguishing the first stars.

\acknowledgements We are grateful to Mark Giroux for calculating the
IGM transfer functions presented in Section 4, and to Claus Leitherer
and his collaborators for the public distribution of their Starburst99
population synthesis code (http://www.stsci.edu/starburst99). Comments
from John Stocke, Nick Gnedin, and Andrew Hamilton improved the
manuscript. We happily acknowledge the anonymous referee who prompted
us to clarify the discrepancy between our emission line predictions and
those of Schaerer (2002), and who provided other useful comments. The
NGST design specifications were taken from the NGST website in April
2002 (http://www.stsci.edu/science/sensitivity). The OPAL project at
Lawrence Livermore National Laboratory is gratefully acknowledged for
their opacity data and interpolation codes
(http://www-phys.llnl.gov/Research/OPAL/opal.html). This work was
partially supported by an astrophysical theory grant to the University
of Colorado by NASA (NAG5-7262).

\begin{table}
\caption[Reactions included in the stellar evolution
models.]{Reactions included in the stellar evolution models.}
\renewcommand{\arraystretch}{1.0}
\vspace{0.05in} \centerline{\label{rxntable}
\begin{tabular}{ll}
\hline \hline
Class & Reaction    \\
\hline
  PPI      &  $^1$H + $^1$H $\rightarrow$ $^2$H + $\nu _e + e^{+}$         \\
            &  $^2$H + $^1$H $\rightarrow$ $^3$He + $\gamma$               \\
            &  $^3$He + $^3$He $\rightarrow$ $^4$He + 2$^1$H               \\
   PPII     &  $^3$He + $^4$He $\rightarrow$ $^7$Be + $\gamma$             \\
            &  $^7$Be + e$^{-}$ $\rightarrow$ $^7$Li + $\nu_e$             \\
            &  $^7$Li + $^1$H $\rightarrow$ 2$^4$He                        \\
   PPIIII   &  $^7$Be + $^1$H $\rightarrow$ $^8$B + $\gamma$               \\
            &  $^8$B  $\rightarrow$ $^8$Be$^*$ + $e^+$ + $\nu _e$          \\
            &  $^8$Be$^*$ $\rightarrow$ 2$^4$He                            \\
   CNO      &  $^{12}$C + $^1$H  $\rightarrow$ $^{13}$N + $\gamma$         \\
            &  $^{13}$N          $\rightarrow$ $^{13}$C + e$^+$ + $\nu_e$  \\
            &  $^{13}$C + $^1$H  $\rightarrow$ $^{14}$N + $\gamma$     \\
            &  $^{14}$N + $^1$H  $\rightarrow$ $^{15}$O + $\gamma$     \\
            &  $^{15}$O          $\rightarrow$ $^{15}$N + e$^+$ + $\nu_e$  \\
            &  $^{15}$N + $^1$H  $\rightarrow$ $^{12}$C + $^4$He   \\
            &  $^{15}$N + $^1$H  $\rightarrow$ $^{16}$O + $\gamma$ \\
            &  $^{16}$O + $^1$H  $\rightarrow$ $^{17}$F + $\gamma$ \\
            &  $^{17}$F          $\rightarrow$ $^{17}$O + e$^+$ + $\nu_e$  \\
            &  $^{17}$O + $^1$H  $\rightarrow$ $^{14}$N + $^4$He   \\
            &  $^{17}$O + $^1$H  $\rightarrow$ $^{18}$O + e$^+$ + $\nu_e$ \\
   3$\alpha$&  $^4$He + $^4$He $\rightarrow$ $^8$Be$^*$                  \\
            &  $^8$Be$^*$ $\rightarrow$ 2$^4$He       \\
            &  $^8$Be$^*$ + $^4$He $\rightarrow$ $^{12}$C + $\gamma$    \\
$\alpha$ capture & $^{12}$C + $^4$He $\rightarrow$ $^{16}$O + $\gamma$    \\
                 & $^{16}$O + $^4$He $\rightarrow$ $^{20}$Ne + $\gamma$  \\
                 & $^{20}$Ne + $^4$He $\rightarrow$ $^{24}$Mg + $\gamma$  \\
                 & $^{14}$N + $^4$He   $\rightarrow$ $^{18}$F + $\gamma$  \\
                 & $^{18}$O + $^4$He $\rightarrow$ $^{22}$Ne + $\gamma$   \\
                 & $^{22}$Ne + $^4$He $\rightarrow$ $^{25}$Mg + n  \\
 \hline
\end{tabular}}
\vspace{0.2in}
\end{table}

\begin{table}
\caption[Pop III H-burning lifetimes]{Pop III H-burning lifetimes}
\renewcommand{\arraystretch}{1.0}
\vspace{0.05in} \centerline{\label{lifetable}
\begin{tabular}{cc}
\hline \hline
Mass (\Msun) & $\tau_{H}$ (yr)   \\
\hline
 1 &   5.87$\times$10$^9$ \\
 3 &   2.01$\times$10$^8$ \\
 5 &   6.03$\times$10$^7$ \\
 7 &   3.11$\times$10$^7$ \\
 10 &  1.76$\times$10$^7$ \\
 12 &  1.40$\times$10$^7$ \\
 15 &  1.11$\times$10$^7$ \\
 20 &  8.47$\times$10$^6$ \\
 30 &  5.86$\times$10$^6$ \\
 50 &  4.09$\times$10$^6$ \\
 70 &  3.42$\times$10$^6$ \\
 100 & 2.93$\times$10$^6$ \\
 \hline
\end{tabular}}
\vspace{0.2in}
\end{table}

\pagebreak

\begin{figure*}
\plotone{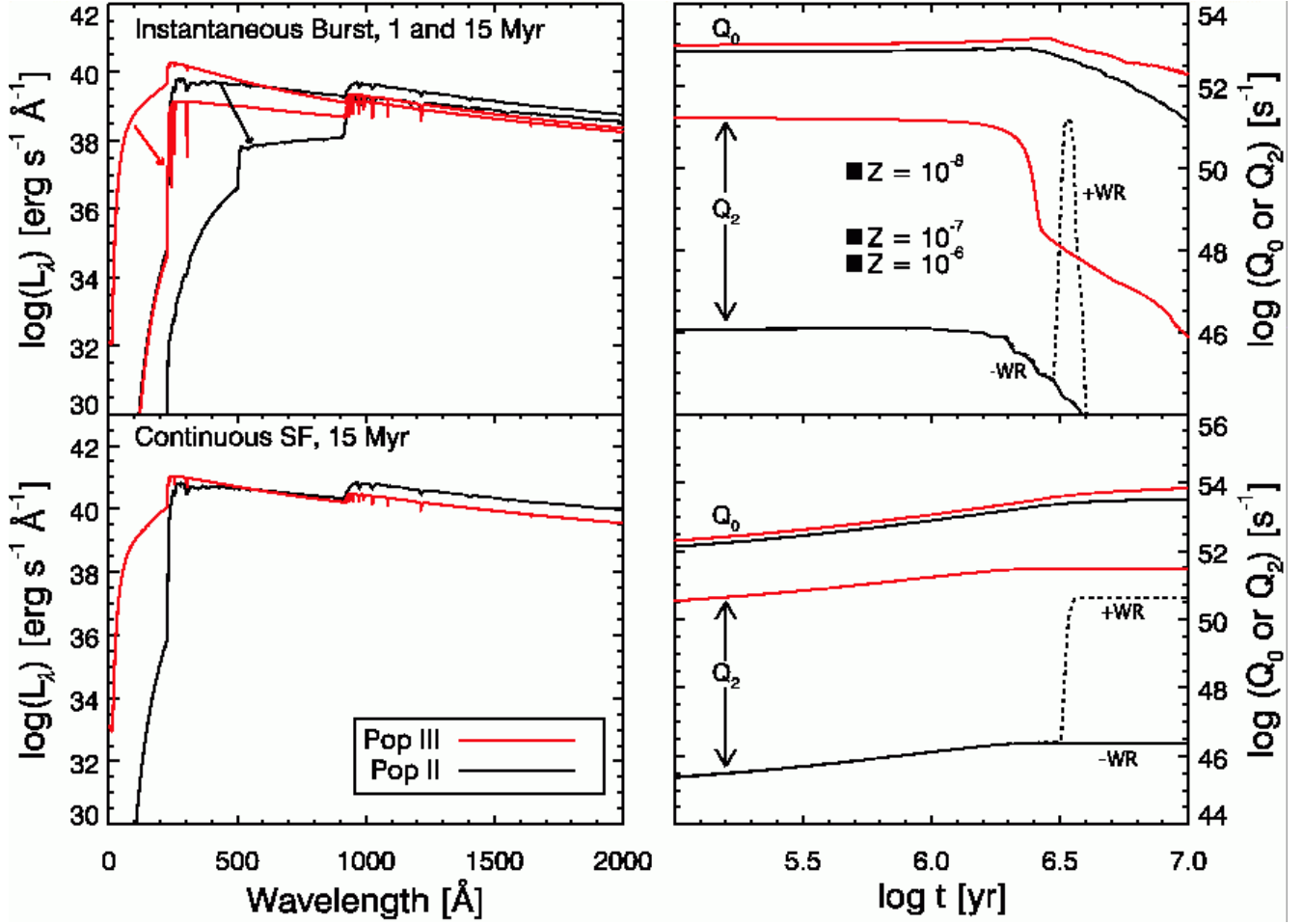}
\figcaption[Comparison of evolving spectra and Q for Pop II and Pop III
clusters.]{Upper left: composite spectra for Pop II (black) and Pop III
(red) clusters ($M = 10^6$ \Msun\ in stars) at 1 and 15 Myr. At 15 Myr
the Pop II spectrum has faded in H~I ionization, but the Pop III
cluster is still a significant source of H~I ionization owing to the
presence of stars with $M = 10 - 15$ \Msun\ and $T_{\rm eff} > 50,000$
K. No nebular emission is included here. Lower left:  example spectra
for the continuous star formation case at 15 Myr.  Upper and lower
right: cluster $Q_0$ and $Q_2$ corresponding to the
instantaneous/continuous cases at left. H~I ionization from the Pop III
cluster is 60\% stronger than Pop II. Pop III emits $10^5$ times more
He~II ionizing photons than Pop II with the same IMF, total mass, 
and excluding WR stars. If WR stars are included, this gain in He II
ionization is smaller but still substantial. This key result has
potentially large effects on the IGM (see Paper II). In the upper right
panel we mark with filled squares the instantaneous values of $Q_2$
from the zero-age main sequences with $Z_C = $ 10$^{-8}$, 10$^{-7}$,
and 10$^{-6}$ (plotted at an arbitrary time). These points show the
sharp decline in He~II ionizing photon production when small abundances
of $^{12}$C are included. The right panels also show the
Starburst99 results with WR stars (dotted lines), for comparison. See
\S~4.2 for discussion. \label{fig:spec+qfig}}
\end{figure*}

\begin{figure*}
\plotone{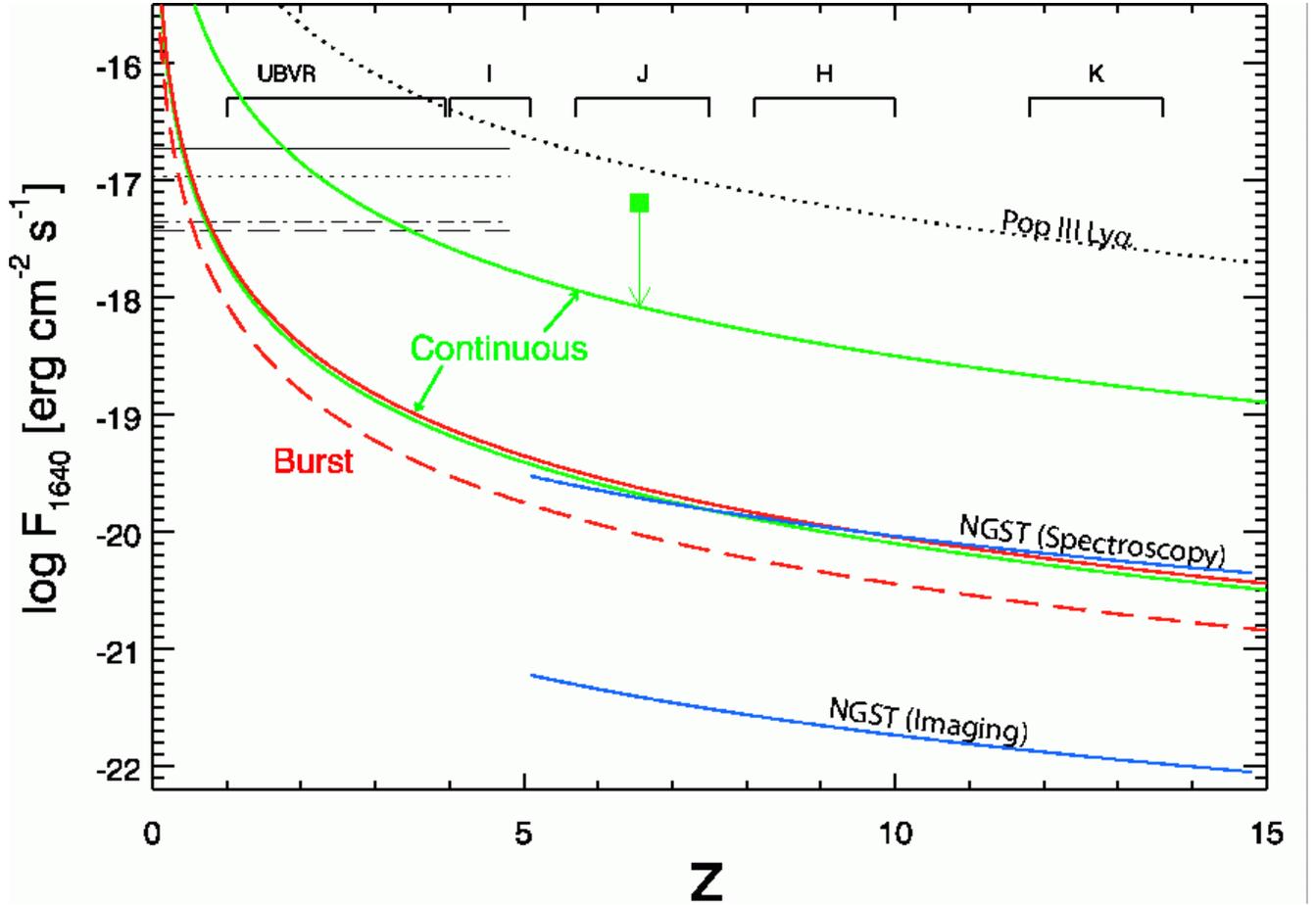}
\figcaption[Predictions for nebular emission line fluxes from
metal-free stars.]{Here we show the predicted fluxes for
Ly$\alpha$ and He~II $\lambda$1640 from the Pop III cluster.
Ly$\alpha$ emission may serve as ``signposts'' pointing the way to
metal-free star formation, which could then be confirmed by the
detection of He~II lines. The green square marks the observed
Ly$\alpha$ flux detected from the $z = 6.56$ galaxy found by Hu et
al. (2002), and the vertical arrow connects this point to the
predicted He~II flux curve if this galaxy harbors an continuous
Pop III starburst with SFR = 40 \Msun\ yr$^{-1}$. The lower green
curve corresponds to SFR = 1 \Msun\ yr$^{-1}$. The red curves mark
the He~II $\lambda$1640 flux for an instantaneous burst at 0 and 1
Myr (solid and dashed, respectively). The blue curves plot the
imaging and spectroscopic sensitivities for NGST. At the top are
wavelength ranges of the common photometric bandpasses. The
horizontal lines extending to $z = 5$ represent the sensitivity
limits of recent emission-line surveys of galaxies: from top to
bottom, Rhoads et al. (1999), Hu et al. (1999; imaging data),
Stern et al. (1999), and Hu et al. (1999; spectroscopic data).
\label{fig:obs}}\vspace{0.05in}\end{figure*}

\begin{figure*}[t]
\epsfxsize=\hsize{\epsfbox{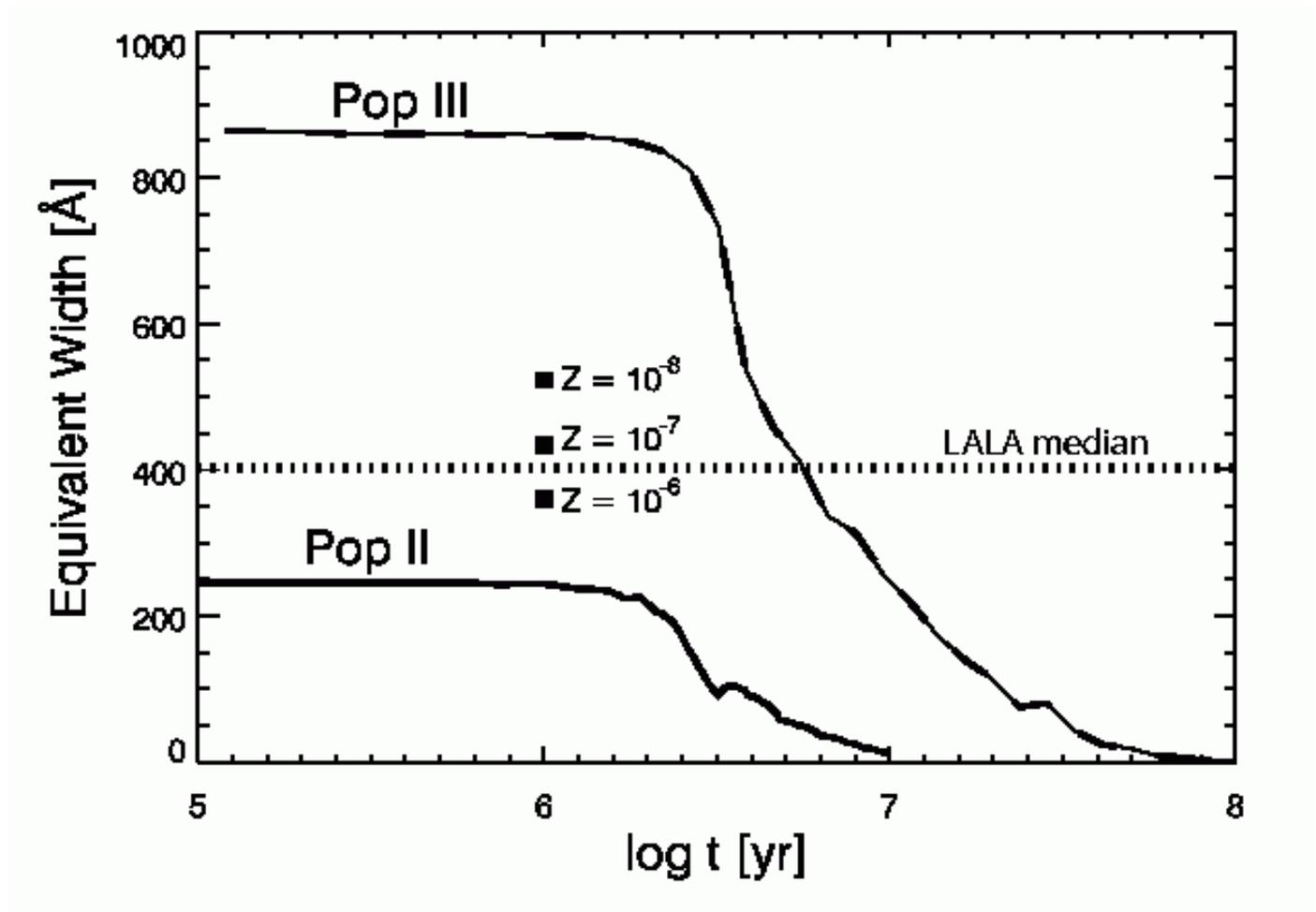}} \figcaption[Comparison of Pop II
and Pop III Ly$\alpha$ equivalent widths.]{Predicted equivalent width,
$W_{\lambda}$, in the Ly$\alpha$ emission line for the Pop III
synthetic cluster (a Salpeter IMF up to 100 \Msun). The Pop III
equivalent width is unusually large, owing to the harder Pop III
spectra. This horizontal dashed line represents the 80 \AA\ detection
limit of the Large Area Ly$\alpha$ survey (Malhotra \& Rhoads 2002).
The horizontal dotted line marks 400 \AA, the median equivalent width
for their sample of 157 galaxies. The filled squares mark the
instantaneous $W_{\rm Ly\alpha}$ from the low-metallicity ZAMS, placed
arbitrarily at $t = 10^6$ yr for clarity. Pop III clusters may be
identified by their large Ly$\alpha$ equivalent widths before their
continuum spectra or broadband colors are measured. However, we note
that at Pop II metallicities, a top-heavy IMF or one with stars more
massive than 100 \Msun\ can mimic this effect.
\label{fig:eqw}}\vspace{0.05in}\end{figure*}

\begin{figure*}
\centerline{\epsfxsize=0.90\hsize{\epsfbox{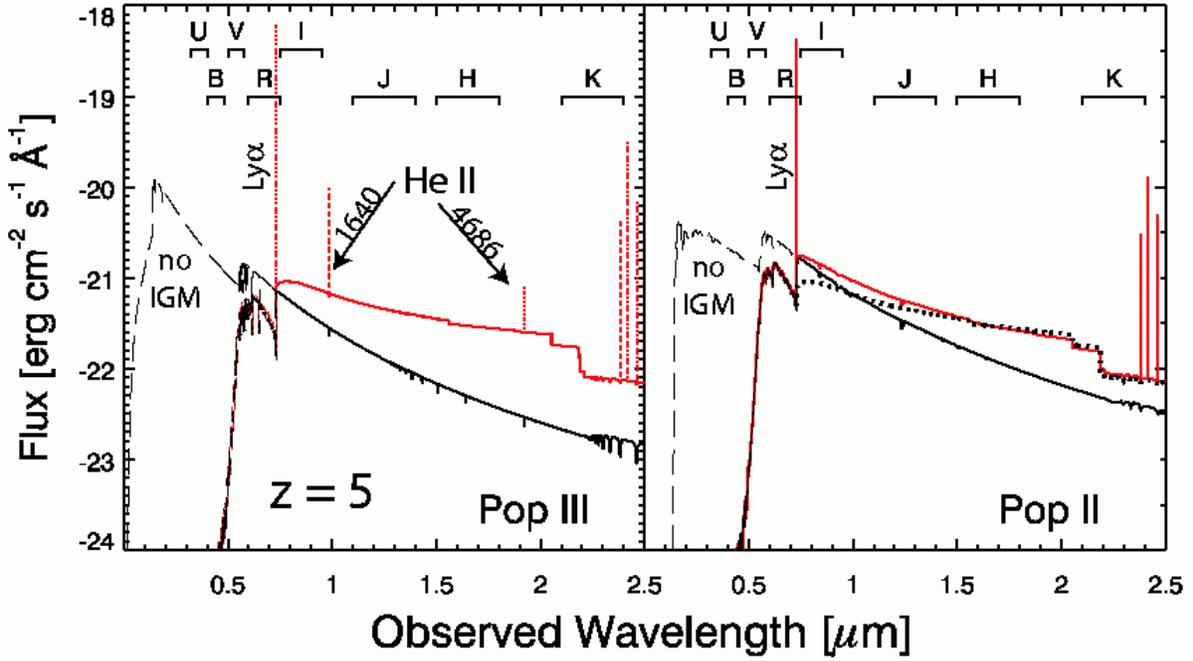}}}
\figcaption[Example observed spectra of Pop II and Pop III
clusters at $z=5$.]{Example observed spectra, at time $t = 0$, for
the Pop III (left) and Pop II (right) cluster at $z = 5$.
Shortward of the rest-frame Lyman limit, the spectra are marked
with long-dashed lines. The black lines mark the intrinsic stellar
spectra, and the red lines include the nebular continuum emission.
Labeled along the top are the common optical and infrared
photometric bandpasses. The dotted line in the right panel repeats
the Pop III spectrum at left for comparison. The spectral regions
where Pop III luminosity exceeds that of Pop II lie in the range
heavily attenuated by the IGM ($\lambda < 1216(1+z)$ \AA), which
is represented here by a mean transmission for 1000 sightlines
simulated by a Monte Carlo model of the IGM (Fardal, Giroux, \&
Shull 1998). Note the absence of the He~II recombination lines in
the Pop II spectrum and the similarity between the observed
spectra when nebular continuum emission is included.
\label{fig:spec+emiss}}\vspace{0.05in}\end{figure*}

\begin{figure*}[b]
\epsfxsize=\hsize{\epsfbox{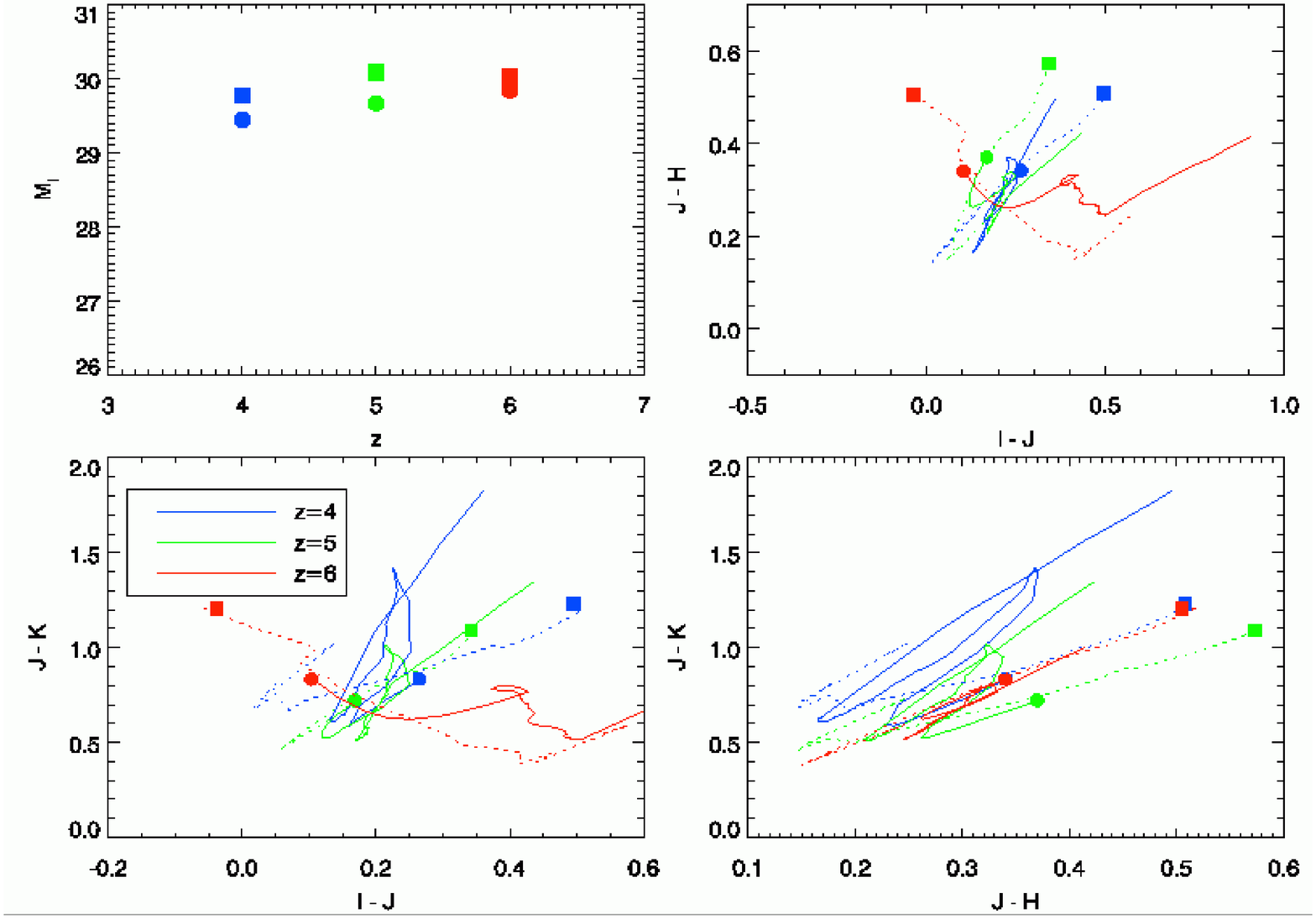}}
\figcaption[Predicted broadband colors of Pop II and Pop III
clusters for $z=4-6$.]{The important broadband color diagnostics
for the Pop III cluster (dotted lines and filled squares),
compared to Pop II (solid lines and filled circles) from the ZAMS
to 10 Myr in age, at redshifts $z$ = 4, 5, and 6 (blue, green, and
red, respectively). The tracks start at the filled symbols. Pop
III clusters are intrinsically fainter and bluer in the infrared
bands, but the influence of nebular continuum emission for $t \leq
10$ Myr hide this difference and give the Pop III and Pop II
clusters similar color signatures. Thus, the recombination lines
of H~I (Ly$\alpha$) and He~II ($\lambda$1640) may prove to be the
best means of distinguishing Pop III stars.
\label{fig:colors}}\vspace{0.05in}
\end{figure*}


\begin{references}

\reference{aller} Aller, L. H., 1984, Physics of Thermal Gaseous
Nebulae (Dordrecht:~Kluwer)

\reference{beck} Becker, R. H., et al. 2001, AJ, 122, 2850


\reference{bromm} Bromm, V., Coppi, P. S., \& Larson R. B. 2000,
ApJ, 527, 5

\reference{bromm} Bromm, V., Kudritzki, R.-P., \& Loeb, A. 2001, ApJ,
552, 464

\reference{cct}Castellani, V., Chieffi, A., \& Tornambe, A. 1983, \apj,
            272, 249

\reference{rates} Caughlan, G. R., \& Fowler, W. A. 1988, Atomic
        Data and Nuclear Data Tables, 40, 283

\reference{cf} Charlot, S., \& Fall, S. M. 1993, ApJ, 415, 580

\reference{d} Djorgovski, S. G., Castro, S. M., Stern, D., \&
Mahabal, A. 2001,  ApJL, 560, L5

\reference{dove} Dove, J. B., Shull, J. M., \& Ferrara, A. 2000,
ApJ, 531, 846

\reference{eleid} El Eid, M. F., Fricke, K. J., \& Ober, W. W.
1983, A\&A, 119, 54

\reference{fan} Fan, X., et al. 2001, AJ, 123, 1247

\reference{fgs} Fardal, M., Giroux, M. L., \& Shull, J. M. 1998,
            AJ, 115, 2206

\reference{go} Gnedin, N. Y., \& Ostriker, J. P. 1997, ApJ, 486,
581

\reference{zh} Haiman, Z. 2002, ApJ, 576, L1

\reference{ht99} Hix, W. R., \& Thielemann, F.-K. 1999, Journal of
    Computational and Applied Mathematics, 99, 321

\reference{hu} Hu, E. M., McMahon, R. G., \& Cowie, L. L. 1999,
ApJ, 522, L9

\reference{hu} Hu, E. M., et al. 2002, ApJ, in press
(astro-ph/0203091)

\reference{tl} Hubeny, I., \& Lanz, T. 1995, ApJ, 439, 875

\reference{opal} Iglesias, C. A., \& Rogers, F. J., 1996, ApJ,
464, 943

\reference{kipp} Kippenhahn, R., Weigert, A., \& Hofmeister, E.
1967, Methods of Computational Physics, 7, 129

\reference{klapp} Klapp, J. 1983, A\&SS, 93, 313

\reference{gp1} Kriss, G., Shull, J. M., Oegerle, W. R., Zheng,
W., Davidsen, A. F., Songaila, A., Tumlinson, J., et al. 2001,
Science, 293, 1112

\reference{kud} Kudritzki, R.-P. 2000, in The First Stars, ESO
Astrophys. Symp., ed. A. Weiss, T. Abel, \& V. Hill (Berlin:
Springer), 127

\reference{kud} Kudritzki, R.-P. 2002, \apj, 577, 389

\reference{lanz} Lanzetta, K. M., et al. 2002, \apj, in press
(astro-ph/0111129)

\reference{lr} Loeb, A.,~\& Rybicki, G.~B. 1999, \apj, 524, 527

\reference{mm} Maeder, A., \& Meynet, G. 1987, A\&A, 182, 243

\reference{mr} Malhotra, S., \& Rhoads, J. E. 2002, ApJ, 565, L71

\reference{mar} Marigo, P., Girardi, L., Chiosi, C., \& Wood, P.
R. 2000, A\&A, 371, 152

\reference{jme} Miralda-Escud\'{e}, J. 1998, ApJ, 501, 15

\reference{agn2} Osterbrock, D. E., 1989, Astrophysics of Gaseous
Nebulae and Active Galactic Nuclei (Mill Valley: University
Science Books)

\reference{max} Pettini, M., Rix, S. A., Steidel, C. C.,
Adelberger, K. L.,  Hunt, M. P., \& Shapley, A. E. 2002 ApJ,
submitted, astro-ph/0110637

\reference{nr} Press, W. H., Flannery, B. P., Teukolsky, S. A., \&
Vetterling, W. T. 1986, Numerical Recipes: The Art of Scientific
Computing (Cambridge: Cambridge University Press)

\reference{r} Rhoads, J. E., Malhotra, S., Dey, A., Stern, D.,
Spinrad, H., \& Jannuzi, B. T. 2000, ApJ, 545, 85

\reference{mr} Ricotti, M., Gnedin, N. Y., \& Shull, J. M. 2002,
ApJ, 575, in press

\reference{mf} Schaerer, D., 2002, A\&A, in press

\reference{sv} Schaerer, D., \& Vacca, W. D. 1998, ApJ, 497, 618

\reference{sch} Schaller, G., Schaerer, D., Meynet, G., \& Maeder,
A., 1992, A\&AS, 96, 269

\reference{s} Schneider, D. P., et al. 2002, AJ, 123, 567

\reference{st} Schramm, D. N., \& Turner, M. S. 1998, Rev Mod
Phys, 70, 303

\reference{he} Seaton, M. J. 1978, MNRAS, 185, 5

\reference{gp} Shull, J. M., Tumlinson, J., et al. 2002, ApJ, in
preparation

\reference{Siess} Siess, L., Livio, M., \& Lattanzio, J. 2002,
ApJ, in press

\reference{ppim} Spitzer, L. 1978, Physical Processes in the
Interstellar Medium (New York: Wiley)

\reference{LBG} Steidel, C. C., et al. 1999, ApJ, 519, 1

\reference{T} Telfer, R. C., Zheng, W., Kriss, G. A., \& Davidsen,
A. F. 2002, ApJ, 465, 773

\reference{fesc} Tumlinson, J., Giroux, M. L., Shull, J. M., \&
Stocke, J. T. 1999, AJ, 118, 2148

\reference{ts} Tumlinson, J., \& Shull, J. M. 2000, ApJ, 528, L65
(TS)

\reference{tgs} Tumlinson, J., Giroux, M. L., \& Shull J. M. 2001,
ApJ, 550, L1 (TGS)

\reference{vts} Venkatesan, A., Tumlinson, J., \& Shull, J. M.
    2002, submitted (Paper II)

\reference{Z} Zheng, W., Kriss, G. A., Telfer, R. C., Grimes, J.
P., \& Davidsen, A. F. 1997, ApJ, 475, 469

\end{references}
\end{document}